\documentclass[aip,cha,amsmath, amssymb,reprint]{revtex4-1}
\usepackage{graphicx}
\usepackage{bm}
\usepackage[caption=false]{subfig}
\usepackage{dcolumn}

\begin{document}

\title{A continuous-time persistent random walk model for flocking}
\author{Daniel Escaff}
\email{descaff@miuandes.cl}
\affiliation{Complex Systems Group, Facultad de Ingenier\'{\i}a y Ciencias Aplicadas,
Universidad de los Andes, Monse\~nor Alvaro del Portillo 12455, Las Condes, Santiago, Chile}
\author{Ra\'ul Toral}
\affiliation{IFISC (Instituto de F{\'\i}sica Interdisciplinar y Sistemas Complejos), Universitat de les Illes Balears-CSIC, 07122-Palma de Mallorca, Spain}
\author{Christian Van den Broeck}
\affiliation{Hasselt University, B-3500 Hasselt, Belgium and Stellenbosch Institute for Advanced Studies, Matieland 7602, South Africa}
\author{Katja Lindenberg}
\affiliation{Department of Chemistry and Biochemistry and BioCircuits Institute, University of California San Diego, La Jolla,
California 92093-0340, USA}
\date{\today}

\begin{abstract}
A classical random walker is characterized by a random position and velocity. This sort of random walk was originally proposed by Einstein to model Brownian motion and to demonstrate the existence of atoms and molecules. Such a walker represents an inanimate  particle driven by environmental fluctuations. On the other hand, there are many examples of so-called ``persistent random walkers'', including self-propelled particles that are able to move with almost constant speed while randomly changing their direction of motion. Examples include living entities (ranging from flagellated unicellular organisms to complex animals such as birds and fish), as well as synthetic materials. Here we discuss such persistent non-interacting random walkers as a model for active particles. We also present a model that includes interactions among particles, leading to a transition to flocking, that is, to a net flux where the majority of the particles move in the same direction. Moreover, the model exhibits secondary transitions that lead to clustering and more complex spatially structured states of flocking. We analyze all these transitions in terms of bifurcations using a number of mean field strategies (all to all interaction and advection-reaction equations for the spatially structured states), and compare these results with direct numerical simulations of ensembles of these interacting active particles. \end{abstract}
\maketitle
\begin{quotation}

Interacting self-propelled particles have the potential to exhibit a number of self-coordinated motions. Nature offers many examples surprising for their beauty, such as flocking birds or swarming fish. The keys to understanding the emergence of such collective behaviors are two: the motion of the self-propelled entities themselves, and the interaction that leads to the coordination. In this work we present a mathematical model for the sort of self-propelled particles that under appropriate conditions are capable of collective motions. This model deepens our understanding of the emergence of collective motion in terms of the theoretical framework provided by nonequilibrium statistical mechanics and nonlinear physics.

\end{quotation}

\maketitle

\section{Introduction}

Brownian motion is one of the main paradigms of stochastic processes in equilibrium statistical physics. Although initially Robert Brown (after whom Brownian motion is named) speculated that there was some remaining life in the pollen grains that he studied, he later observed the same type of motion in dust particles. Einstein instead interpreted this random motion as the result of thermal fluctuations induced by the presence of atoms and molecules colliding with pollen grains or dust particles~\cite{Einstein}, as described by kinetic theory.

Einstein's random walker represents an inanimate particle driven by environmental fluctuations. There are many examples of non-equilibrium self-propelling units in nature. Examples include motor proteins such as myosin~ \cite{Bausch} and kinesin~\cite{Dogic}, and even simpler plastic spheres in a conducting fluid~\cite{Bartolo}. The most complex examples are probably self-propelling living entities, ranging from simple bacteria~\cite{Kaiser,Kasyap} to more complex animal aggregation behaviors~\cite{Parrish} such as flocking birds or swarming fish~\cite{Popkin}.

From the physical point of view, these self-propelled particles are non-equilibrium entities that are able to move at an almost constant speed in a viscous environment. If they interact, they might exhibit self-organized motions. For example, they may exhibit a net flux, where the majority of the particles move in the same direction, a behavior known as \emph{flocking}. Moreover, they can exhibit more complex spatiotemporal collective motions such as the formation of traveling clusters. In 1995, Vicsek \emph{et al.}~\cite{Vicsek} presented the first theoretical evidence of a transition to flocking, proposing a model that has become a paradigm of active matter. The Vicsek \emph{et al.} model is based on a stochastic dynamics, where each particle moves in two dimensions at a constant speed in a random direction chosen at discrete times. That is, the particles execute a random walk in velocity space and at each velocity move ballistically in position space. The selection of these stochastic directions of motion is determined by the average velocity in a vicinity around each active particle. This dependence models the interactions among particles. As a result of these interactions,  the system exhibits a transition to flocking. A few months after Vicsek's publication, Toner and Tu~\cite{Toner} proposed a continuum hydrodynamic-like model for the transition to flocking. They claimed that their theory describes a large universality class of microscopic rules, including Vicsek \emph{et al.}'s (see Ref.~\cite{Toner2} for an extensive review of the Toner-Tu theory).

In both cases, the lower critical dimension for flocking is two. Later on, Vicsek \emph{et al.} modified the model, and observed the flocking transition in one dimension~ \cite{Vicsek2}. 
In the one-dimensional model the particles do not have a constant speed. That is, fluctuations and interactions affect both the magnitude and the direction of the velocity.
 
Even though in the original work of Vicsek \emph{et al.}~\cite{Vicsek} the transition to flocking appeared to be second order (continuous), Gr\'egoire and Chat\'e showed that this result was a finite size effect~\cite{Chate1}. In fact, they showed that, when larger systems are considered, the transition to flocking becomes discontinuous. In contrast to early work on self-propelled particles, Gr\'egoire and Chat\'e claimed that the most general behavior of active matter is a first order (discontinuous) transition to flocking. Their claim was based on several generalizations of the Vicsek model that include vectorial noise and the effect of cohesion. 

The findings of Gr\'egoire and Chat\'e led to an interesting debate. Vicsek's group argued that the transition of the original Vicsek \emph{et al.} model (with scalar noise in position space, leading to diffusion) is second order for low speed of the active particles~\cite{Vicsek3}. Furthermore, they attributed the discontinuous nature of the transition for high speed to a numerical artifact induced byan artificial interplay of a strong anisotropy in the particle diffusion and the periodic boundary conditions. While for low velocities the self-organized state is characterized by small self-propelled clusters, for high velocities it is characterized by density waves. Boundary conditions quantize the propagation direction of the density waves which, in the opinion of Vicsek \emph{et al.}~\cite{Vicsek3}, makes it impossible to determine the physical nature of the flocking transition for higher velocities of the active particles. In addition, Aldana \emph{et al.}~\cite{Aldana} pointed out that the nature of the transition depends crucially on the way in which noise is introduced into the system. To do this, Aldana \emph{et al.} studied a set of networks that are closely related to the problem of self-propelled particles. As a counterargument, Chat\'e \emph{et al.}~\cite{Chate2} claimed that the low speed limit simply increases the system size at which the transition exhibits the discontinuity. That is, they observed  that the transition to flocking becomes first order even at low velocities provided the system size is increased. 

Most of the above mentioned models for active matter are based on hypothetical interactions that are chosen for the sake of simplicity. 
This is the direction that we will also follow in this work. It is worth mentioning, however, that there are other simple active entities (ranging from bacteria to synthetic active particles) which may exhibit more physically motivated interactions. Along this line, for instance, there is a great deal of work that shows that the flocking transition can be observed in self-propelled rods that interact just due to inelastic 
collisions~\cite{Peruani1,Marchetti,Peruani2}. 

Even though Vicsek types of microscopic rules are simple for numerical simulations, it is quite difficult to obtain conclusive analytic results from them. In one spatial dimension Vicsek \emph{et al.}~\cite{Vicsek2} proposed a hydrodynamic-like theory for flocking. More recently, Solon and Tailleur proposed a new kind of microscopic rule that leads to flocking in a model of active spins~\cite{Solon}. Instead of a constant speed, the particles in the Solon-Tailleur model experience anisotropic diffusion, where the direction of anisotropy is dictated by the spin modified by the interaction with neighboring spins. Then, via a coarse-graining procedure, they obtained a set of partial differential equations that describe the system dynamics.

Here we propose a model for flocking based on a particular random walk paradigm, namely, a continuous-time persistent random walk model. In its continuous version it is related to the telegrapher's equation, and in its discrete version, to Kac's walk~\cite{Kac}. A persistent random walker consists of a particle with a constant speed, but with random changes in its direction of motion (as in the usual model for active particles). The properties of noninteracting persistent random walkers and generalizations thereof have been widely studied~\cite{Katja1,Masoliver,Katja2}. In this article we propose and analyze, both theoretically and numerically, a model for interactions which leads to a flocking transition. For the sake of simplicity, we work in one spatial dimension. In Sec.~\ref{Sec2}, we briefly review the continuous-time persistent random walk with no interactions. In Sec.~\ref{Sec3} we present our new model and derive a set of nonlinear partial differential equations that describe the walk with interactions. In Sec.~\ref{Sec4} we implement a mean field approach for the transition to flocking and we also show that there is no spatial structuring of the flocking state via the classical Turing-type of instability. In Sec.~\ref{Sec5} we carry out a detailed numerical analysis of the model and construct the phase diagram of flocking, showing that the formation of traveling clusters is quite robust. In Sec.~\ref{Sec6} we present an analytic estimation of these traveling clusters, showing that the equations derived in Sec.~\ref{Sec3} are in good agreement with the numerical observations. Finally, in Sec.~\ref{Sec7} we summarize and present concluding remarks.

\section{Brief review of persistent random walk}
\label{Sec2}

In this section we briefly introduce the persistent random walk, with the main intention of establishing notation and context for the next sections. The reader interested in this vast topic may consult the extensive literature that has been written about persistent random walks ~\cite{Kac,Katja1,Masoliver,Katja2,PRW}.

\begin{figure}[ht]
\includegraphics[width =3.0 in]{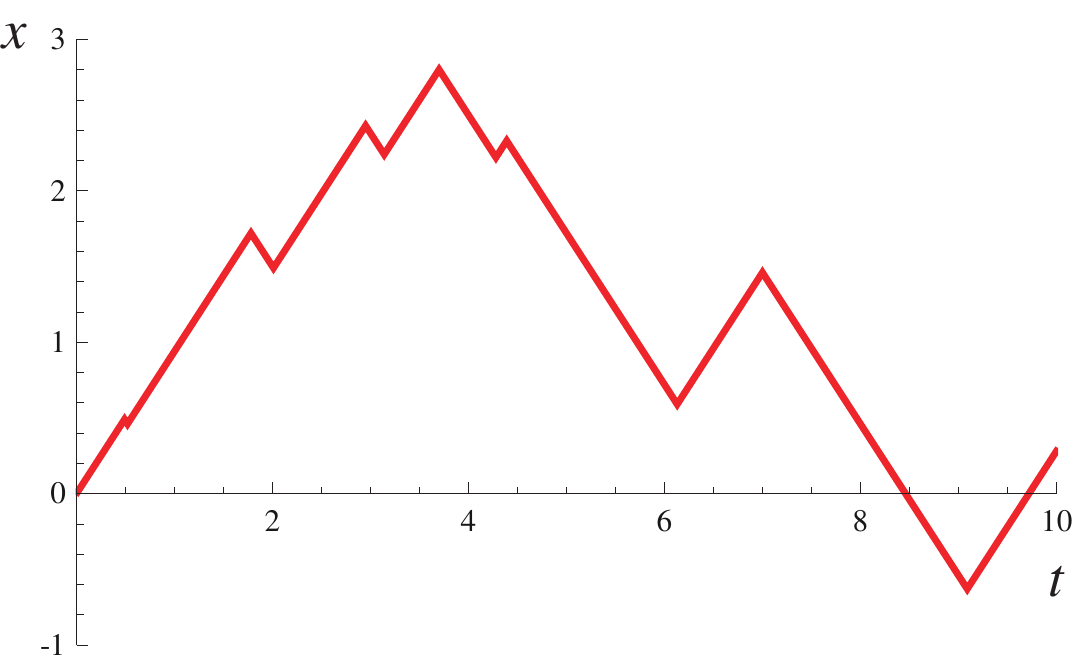}
\caption{Typical trajectory, $x(t)$, of a persistent random walker with $V_0 = 1$ and $\lambda = 1$.} \label{fig1}
\end{figure}

As we mentioned in the introduction, a persistent random walker in one dimension moves at a constant speed, say $V_0$, but can randomly reverse the direction of its motion at a rate $\lambda$. It is thus a spatially extended two-states system: the state of the particle can be characterized by its position $x$, and its direction of motion, that is, direction $+$ (moving to the right) and direction $-$ (moving to the left). Figure~\ref{fig1} shows the typical trajectory of a persistent random walker in which the jumps in the velocity between $V_0$ and $-V_0$ occur at random times that are exponentially distributed. Between these velocity jumps the motion of the walker is ballistic. More precisely, the times between two consecutive jumps obey the waiting time distribution $w(t) = \lambda e^{-\lambda t}$. Hence the mean time between jumps is $\tau = \lambda^{-1}$.

The process can be characterized by two distributions: $\rho_{+}(x,t)$ and $\rho_{-}(x,t)$, where $\rho_{\pm}(x,t)dx$ is the probability of finding the particle at a position within $[x, x + dx]$ and in the state $+$ or $-$ at time $t$. These distributions obey the equations 
\begin{align}
\frac{\partial \rho_{+}}{\partial t} &= - V_0\frac{\partial \rho_{+}}{\partial x} - \lambda (\rho_{+} - \rho_{-}),
\label{EqPRW01}\\
\frac{\partial \rho_{-}}{\partial t} &= V_0\frac{\partial \rho_{-}}{\partial x} + \lambda (\rho_{+} - \rho_{-}).
\label{EqPRW02}
\end{align}

The total probability distribution $\rho(x,t)$ for the particle position $x$ takes the form
\begin{equation}
\rho(x,t) = \rho_{+}(x,t) + \rho_{-}(x,t),\label{PRWprob01}
\end{equation} 
while the flux is given by
\begin{equation}
J(x,t) = V_0\left[\rho_{+}(x,t) - \rho_{-}(x,t)\right].\label{PRWflux01}
\end{equation} 

Equations (\ref{EqPRW01}) and (\ref{EqPRW02}) can be rewritten in terms of $\rho$ and $J$ as
\begin{align}
\frac{\partial \rho}{\partial t} &= - \frac{\partial J}{\partial x},
\label{EqPRW11}\\
\frac{\partial J}{\partial t} &= - V_{0}^{2} \frac{\partial \rho}{\partial x} - 2\lambda J.
\label{EqPRW12}
\end{align}
Equation~(\ref{EqPRW11}) expresses the conservation of the probability, while Eq.~(\ref{EqPRW12}) describes the damping of the flux. If we consider the particle to be confined in a box of size $L$ ($x\in [0,L]$), with periodic or null-flux boundary conditions, then the steady state is 
\begin{equation}
\rho_{st} = 1/L ~~~\text{and}~~~ J_{st} = 0,\label{ST1}
\end{equation} 
that is, a completely uniform distribution in the box, without a preferential direction of motion. 

From Eqs.~(\ref{EqPRW11}) and (\ref{EqPRW12}), we can deduce that the probability $\rho(x,t)$ obeys the telegrapher's equation
\begin{equation}
\frac{\partial^2 \rho}{\partial t^2} + 2\lambda \frac{\partial \rho}{\partial t} - V_{0}^{2} \frac{\partial^2 \rho}{\partial x^2} = 0,
\label{telegrapher}
\end{equation} 
which is perhaps the most common way to describe a persistent random walk. It is a damped wave equation with dispersion relations ($\rho\sim\exp\left(s(k) t + ikx \right)$) of the form
\begin{align}
s_1 (k) &= - \lambda + \sqrt{\lambda^2 - \left(kV_0\right)^2},
\label{dispersion1}\\
s_2 (k) &= - \lambda - \sqrt{\lambda^2 - \left(kV_0\right)^2}.\label{dispersion2}
\end{align}
Note that, for $k=0$, we have $s_1(0) = 0$, which is associated with the conservation of probability, and $s_2(0) = -2 \lambda$, which is associated with the damping of the initial flux. For small $k$ (small gradients),
 \begin{equation*}
s_1 (k) \approx - Dk^2,
\end{equation*}
where $D= V_{0}^{2}\tau/2$ and, as noted earlier,  $\tau = \lambda^{-1}$ is the mean time that a particle spends moving in the same direction. It is interesting  to note the similarity with the swimming diffusivity, $D_{swim}\sim V_{0}^{2}\tau$, obtained in the context of active suspensions~\cite{Burkholder}. Hence, the telegrapher's equation (\ref{telegrapher}) seems to be a good candidate to emulate the properties of active particles in one dimension. Here, the randomization is performed via the jumps in the velocity at rate $\lambda$. 
 
\medskip

\section{The model}
\label{Sec3}

\subsection{Ensemble of $N$ non-interacting active particles}
\label{Sec3A}

We next focus on a ensemble of $N$ non-interacting active particles. At time $t$, there are $N_+(t)$ moving to the right, and $N_-(t)$ moving to the left. The total number of particles is conserved, $N_+(t) + N_-(t) = N$. The state of a particle is characterized by its position and its direction of motion, $+$ or $-$. Therefore, the microscopic state of the system can be described by the set of coordinates
\begin{align}
&\left\{x_{1}^{+}(t), ~ ... ~ , x_{N_+}^{+}(t)\right\},
\nonumber\\
&\left\{x_{1}^{-}(t), ~ ... ~ , x_{N_-}^{-}(t)\right\} ,\nonumber
\end{align}
where $x_i^+(t)$ is the location of the $i$th particle at time $t$ moving right and $x_j^-(t)$ that of the $j$th particle moving left at time $t$. The particles are confined in a one-dimensional box of length $L$, $x_{j}^{\pm}(t)\in \left[ 0, L \right]$ $\forall ~ t$ with $j\in\left\{1, ..., N\right\}$, and with periodic boundary conditions. 

The macroscopic state of the system can be described by the densities of particles in each state,
 \begin{align}
&n_+ (x,t) = \sum_{j=1}^{N_+(t)} \delta \left( x - x_{j}^+(t) \right),\\
&n_- (x,t) = \sum_{j=1}^{N_-(t)} \delta \left( x - x_{j}^-(t) \right).
\label{densitiespm}
\end{align}
Alternatively, we can use the global density and the flux,
\begin{align}
n (x,t)& = n_+ (x,t) + n_- (x,t),
\label{TotalDensity}\\
\mathcal{J} (x,t)& = V_0\left[n_+ (x,t) - n_- (x,t)\right].\label{Flux2}
\end{align}

Note that defining the brackets $\left\langle \ldots\right\rangle$ as the ensemble average, 
 \begin{equation*}
 \left\langle n_{\pm} (x,t) \right\rangle = N \rho_{\pm}(x,t),\label{MVdensitiespm}
\end{equation*}
we have
 \begin{equation*}
 \left\langle n (x,t) \right\rangle = N \rho(x,t) ~~~\text{and}~~~  \left\langle \mathcal{J} (x,t) \right\rangle = N J(x,t).
 \label{MV-DensityFlux}
\end{equation*}

The steady state of a system of non-interacting particles is therefore described by 
 \begin{equation}
\left\langle n_{\pm} (x,t) \right\rangle_{st} = \frac{N}{2L}, ~~ \left\langle n (x,t) \right\rangle_{st} = \frac{N}{L} ~~\text{and}~~  \left\langle \mathcal{J} (x,t) \right\rangle_{st} = 0,
\label{ST2}
\end{equation} 
that is, the global density and flux are $N$ times the density and flux for a single particle.
As expected, an ensemble of non-interacting particles does not exhibit any kind of collective behavior. At the steady state, half of the particles move to the right and the other half move to the left, without any flux.

\medskip

\subsection{Model for interaction}
\label{Sec3B}

In oder to observe the emergence of collective behavior, we must allow the active particles to interact. Let us assume that the particles recognize the densities of particles in each of the two states of motion in a vicinity of range $\sigma$ in each direction, that is,
 \begin{equation}
\mathcal{N}_{\sigma}^{\pm} (x,t) = \frac{1}{2\sigma}\int_{x-\sigma}^{x+\sigma} n_{\pm} (x^{\prime},t) dx^{\prime}.\label{LocalDensity}
\end{equation}
Note that,
 \begin{equation}
\mathcal{N}_{L/2}^{\pm} (x,t) = \frac{N_{\pm}(t)}{L}.
\end{equation}

With an attractive interaction, the probability of a particle to jump from one state of motion to the other will increase with the number of particles that are in the second state. That is, if we denote the rate at which the particle jumps from $\pm$ to $\mp$ as $\lambda\left\{\pm \rightarrow \mp \right\}$, then
\begin{align}\label{rate1}
\lambda\left\{+ \rightarrow - \right\} &= \lambda\left( a\mathcal{N}_{\sigma}^{-} (x,t) \right),\\
\lambda\left\{- \rightarrow + \right\} &= \lambda\left( a\mathcal{N}_{\sigma}^{+} (x,t) \right),\label{rate2}
\end{align}
where $ \lambda\left( z \right)$ is a growing function of its argument $z$ in order to model an attractive interaction between the two states of motion. The parameter $a > 0$ measures the strength of the interaction. 

In order to provide quantitative results, we need a specific model for the growing function $ \lambda\left( z \right)$. Many choices are possible. One could be an exponential to emulate the contact with a thermal bath, as in the Solon-Tailleur model~\cite{Solon}. Of course, there is no reason to assume that this growth will follow a prescription from equilibrium statistical mechanics. For numerical convenience, we have discarded the exponential model. The simplest model for $ \lambda\left( z \right)$ is a linear dependence on $z$. However, the linear model has already been studied in the context of economics by Kirman~\cite{Kirman} with all-to-all interacting agents. He has shown that there is a transition to ordering only for finite numbers $N$ of agents, that is, the ordering is lost in the thermodynamics limit $N\rightarrow \infty$. To avoid these pathological dynamical behaviors, we have chosen a nonlinear model of the form
 \begin{equation}
 \lambda\left( z \right) = A + Bz^{\beta},
 \label{lambmodelNL}
\end{equation}
for which one of us has already shown that the transition to ordering is preserved in the thermodynamic limit with all-to-all interacting agents, the only exception being the linear case $\beta = 1$~\cite{Raul}. Moreover, rescaling the time and the strength of the interaction $a$, we can always set $A=1$ and $B=1$. Here we will restrict ourselves to the quadratic case $\beta = 2$, that is, our working model for $\lambda$ will be
 \begin{equation}
 \lambda\left( z \right) = 1 + z^2.\label{lambmodelPOLI}
\end{equation}
Note that some of us have already analyzed such polynomial rates in the context of all-to-all interactions~\cite{Pinto,Rosas1}, and in a lattice of motionless units~\cite{Rosas2}. Here the consideration of active units introduces new dynamical features.

\medskip

\section{Spatially extended mean field theory for flocking dynamics}
\label{Sec4}

In this section we will derive  a set of partial integro-differential equations that describe the evolution of the macroscopic state of the system. To do this, we will use a mean field strategy similar to the one we used in~\cite{Escaff}, where we dealt with motionless three-state oscillators. Here, since we are dealing with self-propelled units, an advection term appears in the equations. The nonlinearity comes from the interaction, which we refer to as the reaction term in analogy with chemical kinetics. 

Since we are not performing any coarse-graining, the reaction term remains non-local in the macroscopic description. However, we are neglecting the fluctuations. Therefore, the predictions that come from this non-local advection-reaction system should be verified by direct numerical simulations of the microscopic rule that we introduced in the previous section (and that naturally include fluctuations). These comparisons will be made in the following sections.

\medskip

\subsection{Continuos description via advection-reaction equations}
\label{Sec4A}

Note that,
 \begin{equation*}
\left\langle\mathcal{N}_{\sigma}^{\pm} (x,t)\right\rangle = \frac{N}{2\sigma}\int_{x-\sigma}^{x+\sigma} \rho_{\pm} (x^{\prime},t) dx^{\prime}.
\end{equation*}
We introduce the control parameter $\mathcal{C}$ and the interaction ratio $\alpha$, 
 \begin{equation}
\mathcal{C} = \frac{aN}{L} ~~~\text{and}~~~  \alpha =  \frac{2\sigma}{L}.\label{ContPar}
\end{equation}
The control parameter $\mathcal{C}$ may be interpreted as a measure of the intensity of the interaction. We can increase $\mathcal{C}$ in two ways, increasing the coupling strength $a$, or increasing the global density $N/L$. We also define
 \begin{equation}
\nu_{\sigma} \left[\rho_{\pm}(x,t) \right] = \int_{x-\sigma}^{x+\sigma} \rho_{\pm} (x^{\prime},t) dx^{\prime}.\label{NU}
\end{equation}
Then, an ensemble of interacting particles can be described by the non-linear mean field equations
\begin{align}
\frac{\partial \rho_{+}}{\partial t} &= - V_0\frac{\partial \rho_{+}}{\partial x} - \lambda \left( \frac{\mathcal{C}}{\alpha}\nu_{\sigma} \left[\rho_{-} \right]\right)\rho_{+} 
+ \lambda \left( \frac{\mathcal{C}}{\alpha}\nu_{\sigma} \left[\rho_{+} \right]\right)\rho_{-},
\label{MF01}\\
\frac{\partial \rho_{-}}{\partial t} &= V_0\frac{\partial \rho_{-}}{\partial x} + \lambda \left( \frac{\mathcal{C}}{\alpha}\nu_{\sigma} \left[\rho_{-} \right]\right)\rho_{+} 
- \lambda \left( \frac{\mathcal{C}}{\alpha}\nu_{\sigma} \left[\rho_{+} \right]\right)\rho_{-}.
\label{MF02}
\end{align}

For $\lambda$ constant, Eqs.~(\ref{MF01}) and (\ref{MF02}) are equivalent to Eqs.~(\ref{EqPRW01}) and (\ref{EqPRW02}), and predict the absence of collective motion. Note that, independently of the functional form of $\lambda$, Eqs.~(\ref{MF01}) and (\ref{MF02}) always have the solution
 \begin{equation}
\rho_{+} = \rho_{-} = \frac{1}{2L}, \label{Uniform}
\end{equation}
which represents a completely uniform state in space and time, without flux, that is, with no collective behaviors. In fact, it coincides with the steady state of the non-interacting system, e.g. Eq.~(\ref{ST1}) or (\ref{ST2}). However, since the system (\ref{MF01}) and (\ref{MF02}) is nonlinear, the solution (\ref{Uniform}) might destabilize, giving rise to new stable solutions, or may coexist with other stable solutions. These other solutions may represent self-organized states, for instance, a preferential flux  (with both directions equally preferred), or even more complex spatiotemporal structuring. In the next subsection we will explore these possibilities.

\medskip

\subsection{Mean field analysis for the transition to flocking}
\label{Sec4B}

\subsubsection{All-to-all interaction $\sigma=L/2$}
\label{Sec4A1}

We start by analyzing the simplest case of all-to-all interactions, that is, $\sigma=L/2$. Here the system can simply be described by $N_{+}(t)$ and $N_{-}(t)$. Moreover, if we define the probability that a given particle is in state $\pm$ at time $t$, 
 \begin{equation}
P_{\pm}(t) = \int_{0}^{L} \rho_{\pm} (x,t) dx,\label{ProbPM}
\end{equation}
we have
\begin{align}
 \left\langle N_{\pm}(t) \right\rangle &= NP_{\pm}(t),
\nonumber\\
\nu_{L/2} \left[\rho_{\pm}(x,t) \right] &= P_{\pm}(t),
\nonumber
\end{align}
and the normalization condition
 \begin{equation}
P_{+}(t) + P_{-}(t) = 1.\label{Normalizacion}
\end{equation}
Note that consistency between previous limits of integration such as in Eq.~(\ref{NU}) and those of Eq.~(\ref{ProbPM}) implies that $x=L/2$. Since the integral is independent of $x$ the choice does not matter.

Under these conditions, we can integrate Eq.~(\ref{MF01}) over the box $[0,L]$, and use Eq.~(\ref{Normalizacion}), to obtain 
 \begin{equation}
\frac{dP_{+}}{dt} = \lambda \left( \mathcal{C}P_{+} \right)  \left( 1 - P_{+}\right) -  \lambda \left( \mathcal{C}\left( 1 - P_{+}\right) \right)P_{+}.\label{MFglobal}
\end{equation}

Equation~(\ref{MFglobal}) has the fixed point $P_{+} = 1/2$, which represents the homogeneous state (\ref{Uniform}). Self-organization may take place via a destabilization of this solution. This can be studied by the standard linear analysis, that is, with the perturbation 
 \begin{equation}
P_{+} = 1/2 + \varepsilon\exp{\left(s t\right)} .\label{Per}
\end{equation}
Linearizing with respect to the small perturbation parameter $\varepsilon$, we obtain
 \begin{equation}
s = -2\lambda \left( \mathcal{C}/2 \right) + \mathcal{C}\lambda^{\prime} \left( \mathcal{C}/2 \right),\label{lambda}
\end{equation}
where the $^{\prime}$ denotes the derivative with respect to the argument. The symmetric solution $P_{-} = P_{+} = 1/2$ destabilizes when $s > 0$.
The critical point can be calculated specifying the functional form of $\lambda$. For our working model (\ref{lambmodelPOLI}),
 \begin{equation}
 \mathcal{C}_{c} = 2,\label{Ccritico}
\end{equation}
and the system undergoes a supercritical bifurcation (second order transition). For $ \mathcal{C} >  \mathcal{C}_{c}$, $P_{+} = 1/2$ is unstable and two new stable fixed points appear,
 \begin{equation}
P_{\pm} = 1/2 \pm \frac{  \sqrt{\mathcal{C}^2 -  \mathcal{C}_{c}^{2}}}{ \mathcal{C}_{c} ~\mathcal{C}}.\label{NewFixPoint}
\end{equation}

The fixed points (\ref{NewFixPoint}) represent emergence of flocking, that is, the particles self-organize due to the interaction. In order to choose a preferential direction in which the majority moves together, we define the order parameters
\begin{align}
\psi(t) &= \left| \frac{1}{N}\int_{0}^{L} \mathcal{J} (x,t) dx\right| =  \left| \frac{V_0 \left( 2N_+(t) - N\right)}{N}\right|,
\label{OP01}\\
\Psi &= \lim_{T \rightarrow \infty} \frac{1}{T}\int_{0}^{T} \psi(t) dt.
\label{OP02}
\end{align}
With our mean field theory,
\begin{equation}
\Psi_{MF}=
 \left\{
\begin{array}
[c]{ll}
V_0 \sqrt{\mathcal{C}^2 -  \mathcal{C}_{c}^{2}} /\mathcal{C} & \text{ if }  \mathcal{C} > \mathcal{C}_{c} = 2\\
0  & \text{ otherwise }\\
\end{array}
\right. .\label{MFOP2}
\end{equation}

\begin{figure}[ht]
\includegraphics[width =3.0 in]{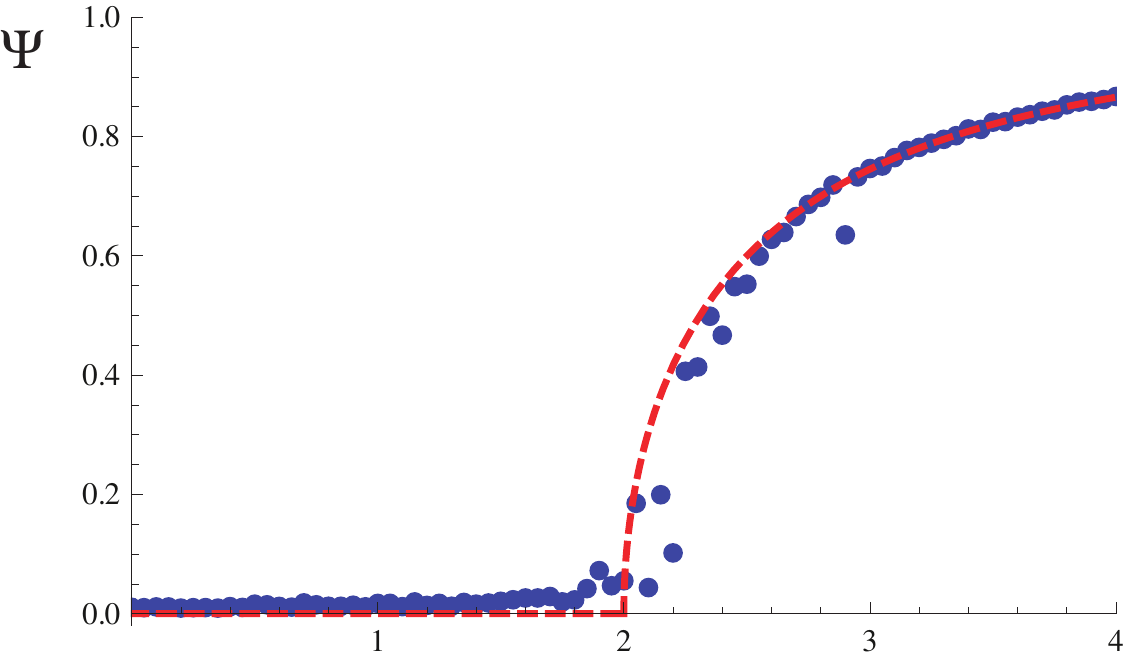}
\caption{Order parameter $\Psi$ versus the control parameter $\mathcal{C}$, for $V_0 = 1$ and $\sigma=L/2$. Dots are the results of a numerical simulation for $N=5000$ with the formula~(\ref{NumOP2}) and with $\Delta t = 10^{-2}$, $T_i = 2$, and $T_f = 20$. The dashed line corresponds to the mean field 
curve~(\ref{MFOP2}).} \label{fig2}
\end{figure}

Figure \ref{fig2} displays the numerical simulation of an ensemble of $N = 5000$ particles, under the effect of global interactions. To estimate the order parameter from the numerical simulations, we have used the prescription 
\begin{equation}
\Psi_{NS} = \frac{\Delta t}{T_f - T_i} \sum_{j=T_i/\Delta t}^{T_f/\Delta t} \left| \frac{V_0 \left( 2N_+( j\Delta t) - N\right)}{N}\right|,\label{NumOP2}
\end{equation}
where $\Delta t$ is the time step of the simulation, $T_i$ is large enough to avoid transient behaviors in the averaging, and $T_f$ is large enough to give a good estimation of the limit in Eq.~(\ref{OP02}).  As can be seen from Fig.~\ref{fig2}, there is good agreement between Eqs.~(\ref{MFOP2}) and (\ref{NumOP2}), although near criticality fluctuations are larger, as expected.

It is worth noting that for this fully connected system, the problem can be solved exactly for finite $N$~\cite{Raul,Pinto,Rosas1}. For instance, in~\cite{Raul}, it has been shown that for the general expression~(\ref{lambmodelNL}), the critical point takes de form 
 \begin{equation*}
 \mathcal{C}_{c} = 2 \left[\frac{A}{B\left(\beta -1 +\beta(3-\beta)N^{-1}\right)}\right]^{1/\beta},
\end{equation*}
which coincides with  expression (\ref{Ccritico}) for $A=B=1$, $\beta=2$, and $N\rightarrow \infty$, as expected.

\subsubsection{Absence of Turing-type instabilities in the case  $\sigma < L/2$}
\label{Sec4A2}

The branches in Eq.~(\ref{MFOP2}) are still valid for the case $\sigma < L/2$. For $ \mathcal{C} > \mathcal{C}_{c}$, they represent a uniform flux, without any spatial structuring. However, in this case these branches might destabilize due to a finite wavelength instability, which leads to a spatial pattering of the flocking state. This is the classical Turing instability, first proposed in the context of reaction-diffusion 
systems~\cite{Turing}. It is worth mentioning that the Turing mechanism has been widely explored for non-local interactions in many contexts such as population dynamics~\cite{Fuentes,Emilio1,Emilio2,Escaff2,Escaff3}, synchronization~\cite{Escaff} and vegetation patterning in arid zones~\cite{Lejeune,Escaff4}, just to mention a few examples. Furthermore, finite wavelength instabilities have also been found in the context of hydrodynamics-like coarse-grained descriptions of active matter \cite{Marchetti2,Marchetti3,Ihle}. For our working model, however, we have not found any Turing-type instability of the uniform states. Below we briefly summarize our results for the advection-reaction 
equations (\ref{MF01}) and (\ref{MF02}).

Let us consider a perturbation in Fourier space for the disordered state (\ref{Uniform}), that is,
 \begin{equation}
\rho_{\pm} = \frac{1}{2L} + \varepsilon_{\pm}\exp{\left(s t + ikx\right)}. \label{Per2}
\end{equation}
Introducing  Eq.~(\ref{Per2}) into Eqs.~(\ref{MF01}) and (\ref{MF02}), and linearizing with respect to $\varepsilon_{\pm}$, we obtain an eigenvalue problem for $s$ which admits the two solutions
\begin{align}
s_1 (k) &= - \Lambda(k) + \sqrt{\Lambda(k)^2 - \left(kV_0\right)^2},
\label{Eigen1}\\
s_2 (k) &= - \Lambda(k) - \sqrt{\Lambda(k)^2 - \left(kV_0\right)^2},\label{Eigen2}
\end{align}
where
 \begin{equation}
 \Lambda(k) = \lambda \left( \mathcal{C}/2 \right) - \frac{\mathcal{C}}{2}\lambda^{\prime} \left( \mathcal{C}/2 \right)  \left\{\frac{\sin k\sigma}{k\sigma} \right\}.\label{Lambda}
 \end{equation}
 
Note that, $s_1 (0) = 0$, which is associated with the conservation of probability. On the other hand, $s_2 (0) = s$, where $s$ is given by Eq.~(\ref{lambda}). Therefore, for $k=0$ the system reproduces the features of the globally coupled ensemble. For $\lambda$ constant, Eqs.~(\ref{Eigen1}) and (\ref{Eigen2}) reduce to Eqs.~(\ref{dispersion1}) and (\ref{dispersion2}). That is, without interactions, 
Eqs.~(\ref{Eigen1}) and (\ref{Eigen2}) correspond to the dispersion relation of the telegrapher's equation. 
 
A Turing-type instability requires that the real part of one of the eigenvalues in (\ref{Eigen1}) and (\ref{Eigen2}) become positive for a finite wavelength (i.e., $k\neq 0$). This occurs when $ \Lambda(k)$ becomes negative. Since $\sin(k\sigma)/k\sigma$ has its maximum at $k=0$, the first mode to become unstable corresponds to $k=0$, with the critical point (\ref{Ccritico}) for the interaction model (\ref{lambmodelPOLI}). Therefore, the instability of the disordered state for $\sigma < L/2$ has the same features as for all-to-all interactions, $\sigma = L/2$. Hence, no Turing mechanism spatial structuring is expected. 
 
Furthermore, we can check the stability of the uniform flocking branches. That is, checking the stability under perturbations of the form
 \begin{equation}
\rho_{\pm} = \frac{Q_{\pm}}{L} + \varepsilon_{\pm}\exp{\left(s t + ikx\right)}, \label{Per3}
\end{equation} 
where 
 \begin{equation*}
Q_{\pm} = \frac{1}{2} \pm \frac{  \sqrt{\mathcal{C}^2 -  \mathcal{C}_{c}^{2}}}{ \mathcal{C}_{c} ~\mathcal{C}}, 
\end{equation*} 
corresponds to spatially uniform flocking, with a net movement to the right (the analysis for flocking to the left is completely equivalent). Note that we have explicitly used the model (\ref{lambmodelPOLI}), and restricted the analysis to $\mathcal{C} >  \mathcal{C}_{c}$. 

In this case, the eigenvalue problem gives us 
\begin{align}
s_1 (k) &= - \bar{\Lambda}(k) + \sqrt{\bar{\Lambda}(k)^2 - \left(kV_0\right)^2 + ikV_0\Delta(k)},
\label{Eigen1f}\\
s_2 (k) &= - \bar{\Lambda}(k) - \sqrt{\bar{\Lambda}(k)^2 - \left(kV_0\right)^2 + ikV_0\Delta(k)},\label{Eigen2f}
\end{align}
where
\begin{align}
\bar{\Lambda}(k) &= \frac{1}{2}\left[ \Lambda_{+}(k) + \Lambda_{-}(k)\right] ,
\label{LambdaBAR}\\
\Delta(k) &=\Lambda_{+}(k) - \Lambda_{-}(k),\label{Delta}
\end{align}
with
 \begin{equation*}
\Lambda_{\pm}(k) = \lambda \left( \mathcal{C}Q_{\mp} \right) - \mathcal{C}Q_{\mp}\lambda^{\prime} \left( \mathcal{C}Q_{\pm} \right)  \left\{\frac{\sin k\sigma}{k\sigma} \right\}.
\end{equation*} 

In this case, the spectra (\ref{Eigen1f}) and (\ref{Eigen2f}) again do not show any positive values in its real parts. Therefore, the Turing mechanism for spatial structuring is, again, absent in the spatially uniform flocking states. However, spatial structuring may appear due to other mechanisms which do not involve a destabilization of the spatially uniform states. In fact, as we will see below, clustering is very often encountered for low $\sigma$.

\section{Numerical observations and phase diagrams for flocking}
\label{Sec5}

We have performed numerical simulations of the stochastic process defined by the rates Eqs.(\ref{rate1}), (\ref{rate2}), and (\ref{lambmodelPOLI}) for different values of the interaction distance $\sigma$.  

\subsection{All-to-all interactions}
We first consider the case of all-to-all interactions where the length $L$ is irrelevant. Recall that in this case the mean-field prediction for the transition point is $\mathcal{C}_c=2$.

\begin{figure}[ht]
\includegraphics[width =0.50\textwidth]{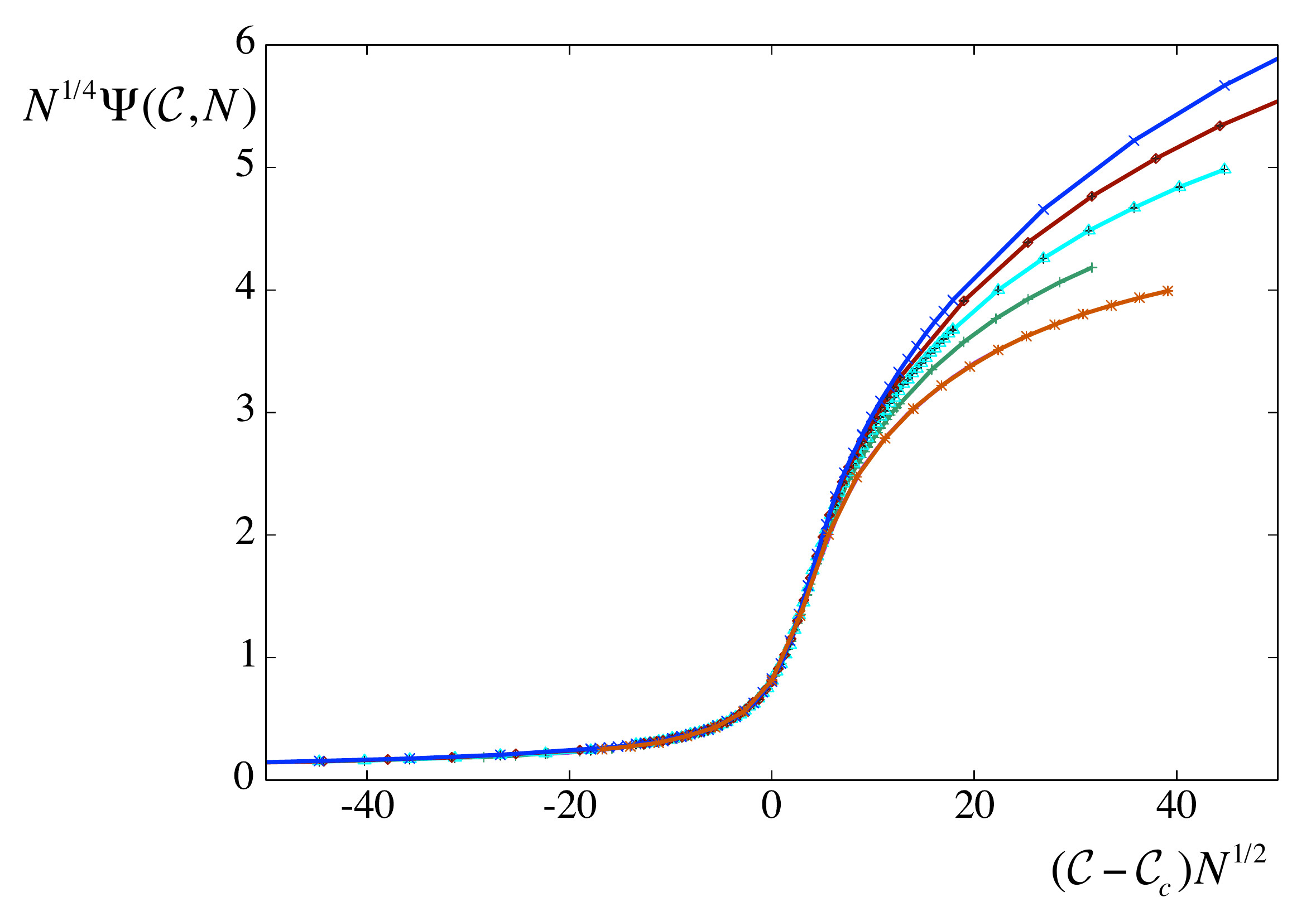}
\caption{Plot of $N^{1/4}\Psi(\mathcal{C},N)$ versus $(\mathcal{C}-\mathcal{C}_c)N^{1/2}$. The data collapse valid in a large interval of the x-coordinate indicates the validity of the finite-size-scaling law using the Ising universality-class critical exponents. The data (from bottom to top at the right of the figure, the lines are a guide to the eye) correspond to $N=500,1000,2000,4000,8000$.} \label{fig:scalpsi}
\end{figure}

\begin{figure}[ht]
\includegraphics[width =0.50\textwidth]{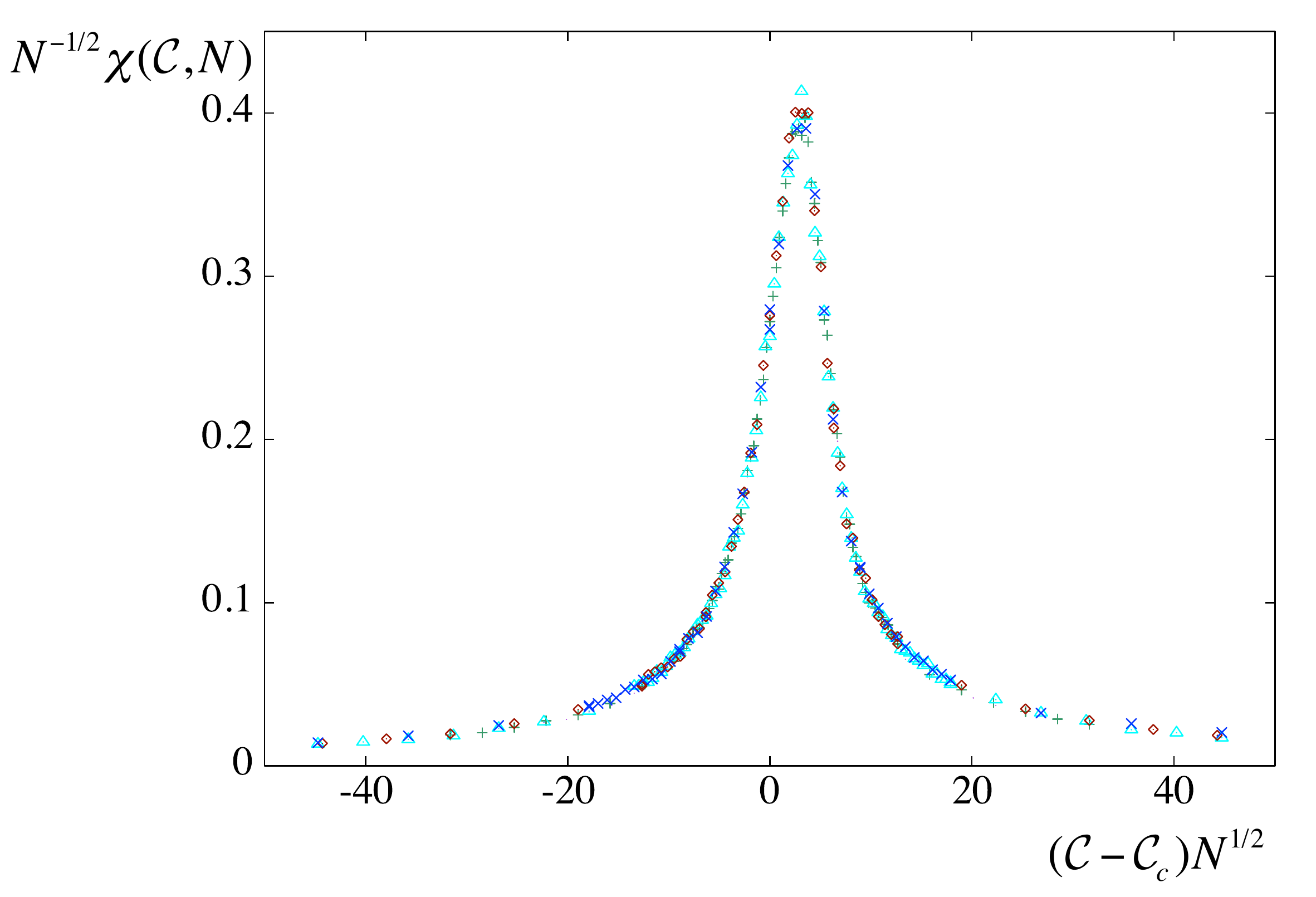}
\caption{Plot of $N^{-1/2}\chi(\mathcal{C},N)$ versus $(\mathcal{C}-\mathcal{C}_c)N^{1/2}$. The data collapse valid in a large interval of the x-coordinate indicates the validity of the finite-size-scaling law using the Ising universality-class critical exponents. The data correspond to $N=500,1000,2000,4000,8000$.} \label{fig:scalchi}
\end{figure}

We have already shown in Fig.~\ref{fig2} that the order parameter $\Psi=\langle \psi\rangle$ obtained from the numerical simulations and the order parameter obtained from the analytic theory agree quite well. Of course, small deviations from the theory are to be expected as perfect agreement should only occur in the thermodynamic limit $N\to\infty$. We have found that the data for different values of $N$ can be accommodated in a finite-size-scaling form $\Psi(\mathcal{C},N)=N^{-A}f_\Psi(\epsilon N^B)$, with $\epsilon=\mathcal{C}-\mathcal{C}_c=\mathcal{C}-2$ and $f_\Psi(x)$ is the scaling function. Evidence for this scaling behavior is shown in Fig.~\ref{fig:scalpsi} using the Ising universality class critical exponents~\cite{Deutsch} $A=1/4$, $B=1/2$. Further evidence that this model in the all-to-all limit belongs to the universality class of the Ising model is given by analyzing the critical behavior of the normalized fluctuations of the order parameter (the ``magnetic susceptibility'' in the Ising model language) $\chi=N\left[\langle \psi^2\rangle -\langle \psi\rangle^2\right]$. In the thermodynamic limit it diverges at the critical point as $\chi(\mathcal{C})\sim |\mathcal{C}-\mathcal{C}_c|^{-\gamma}$, with a critical exponent $\gamma=1$. Finite-size-scaling theory predicts that data for different system sizes should behave as $\chi(\mathcal{C},N)=N^{C}f_\chi(\epsilon N^B)$, with $\epsilon=\mathcal{C}-\mathcal{C}_c=\mathcal{C}-2$ and $f_\chi(x)$ is the scaling function. Evidence for this scaling behavior is shown in Fig.~\ref{fig:scalchi} again using the Ising universality class critical exponents $C=1/2$,  $B=1/2$.

\subsection{Finite-range interactions}

We now consider the case of a finite range of interaction $\sigma$. In the numerical simulations we have taken $L=1$, a constant number of particles $N=10^3$ and varied $\sigma$ in the interval $\sigma\in(0.05,0.5)$ for different values of the coupling constant $\mathcal{C}$ \cite{morework}. The limiting case $\sigma=0.5$ coincides with the all-to-all situation discussed in the previous subsection. It is remarkable that the order parameter $\Psi$ and its normalized fluctuations $\chi$ are independent of $\sigma$ for all values $\sigma\gtrsim 0.3$. As $\sigma$ decreases the order parameter starts to depend on $\mathcal{C}$ and the transition becomes discontinuous at a transition value $\mathcal{C}^*(\sigma)<2$. The normalized fluctuations $\chi$ are displayed in Fig.\ref{fig:chi}.

\begin{figure}[ht]
\includegraphics[width =0.50\textwidth]{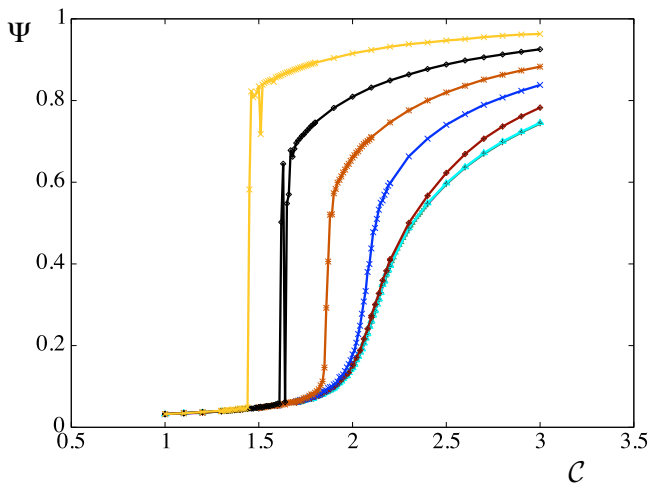}
\caption{Plot of $\Psi(\mathcal{C},N)$ versus $\mathcal{C}$ for system size $N=1000$, physical extension $L=1$ and different values of the interaction length $\sigma=0.05,0.10,0.15,0.2,0.25,0.3,0.4,0.5$ (from left to right in the figure). Note that the data for $\sigma=0.3,0.4,0.5$ collapse onto the same curve. The jumps between the upper and lower branches at the small values of 
$\sigma$ are an indication of the first-order nature of the transition.} \label{fig:psi}
\end{figure}

\begin{figure}[ht]
\includegraphics[width =0.50\textwidth]{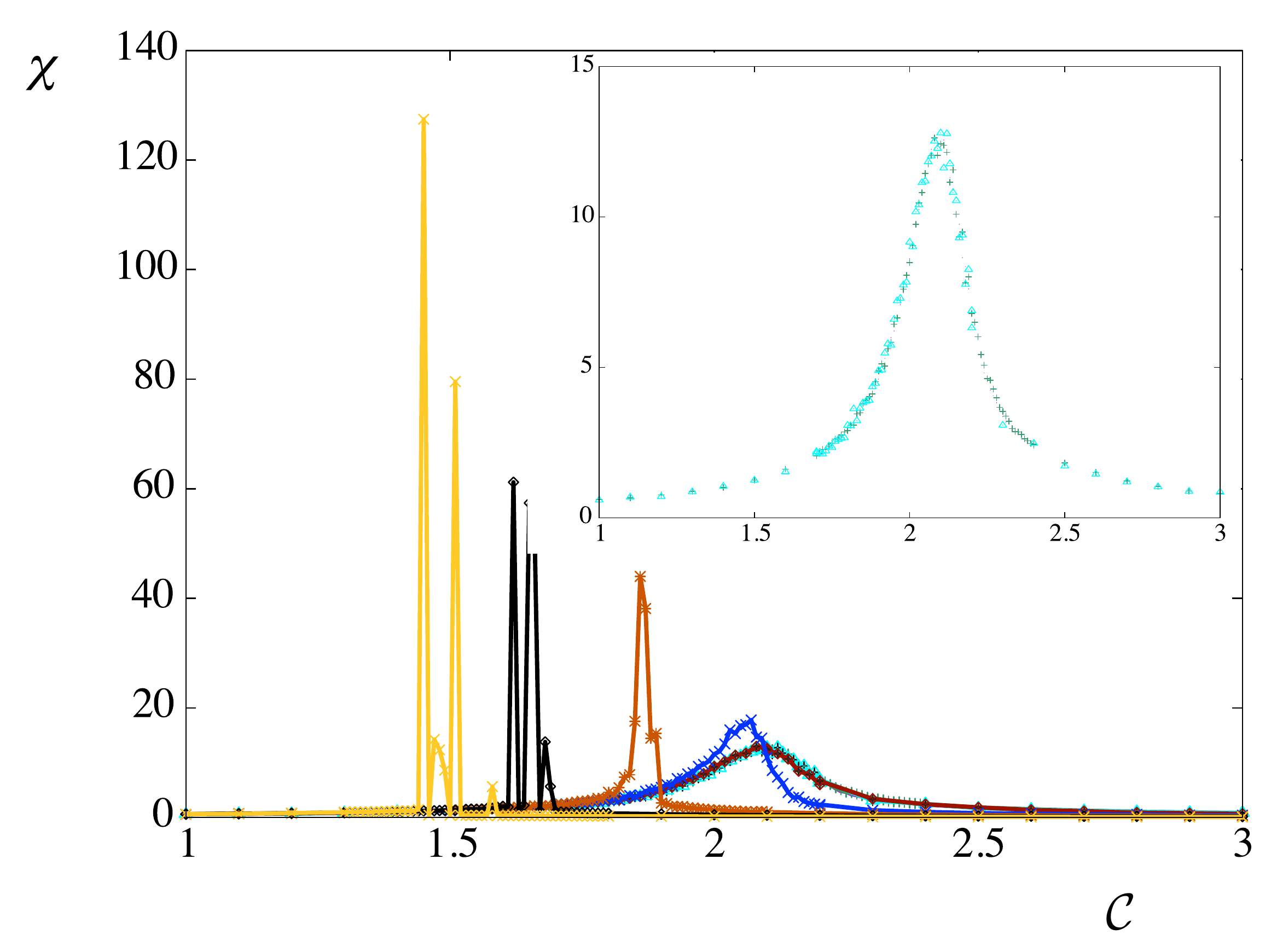}
\caption{Plot of the normalized fluctuation $\chi(\mathcal{C},N)$ versus $\mathcal{C}$ for system size $N=1000$, physical extension $L=1$ and (from left to right in the figure) the same values of the interaction length $\sigma$ used in Fig.\ref{fig:psi}. Note that the data for $\sigma=0.3,0.4,0.5$ collapse to the same curve, as detailed in the insert.} \label{fig:chi}
\end{figure}

\begin{figure}[ht]
\includegraphics[width =0.50\textwidth]{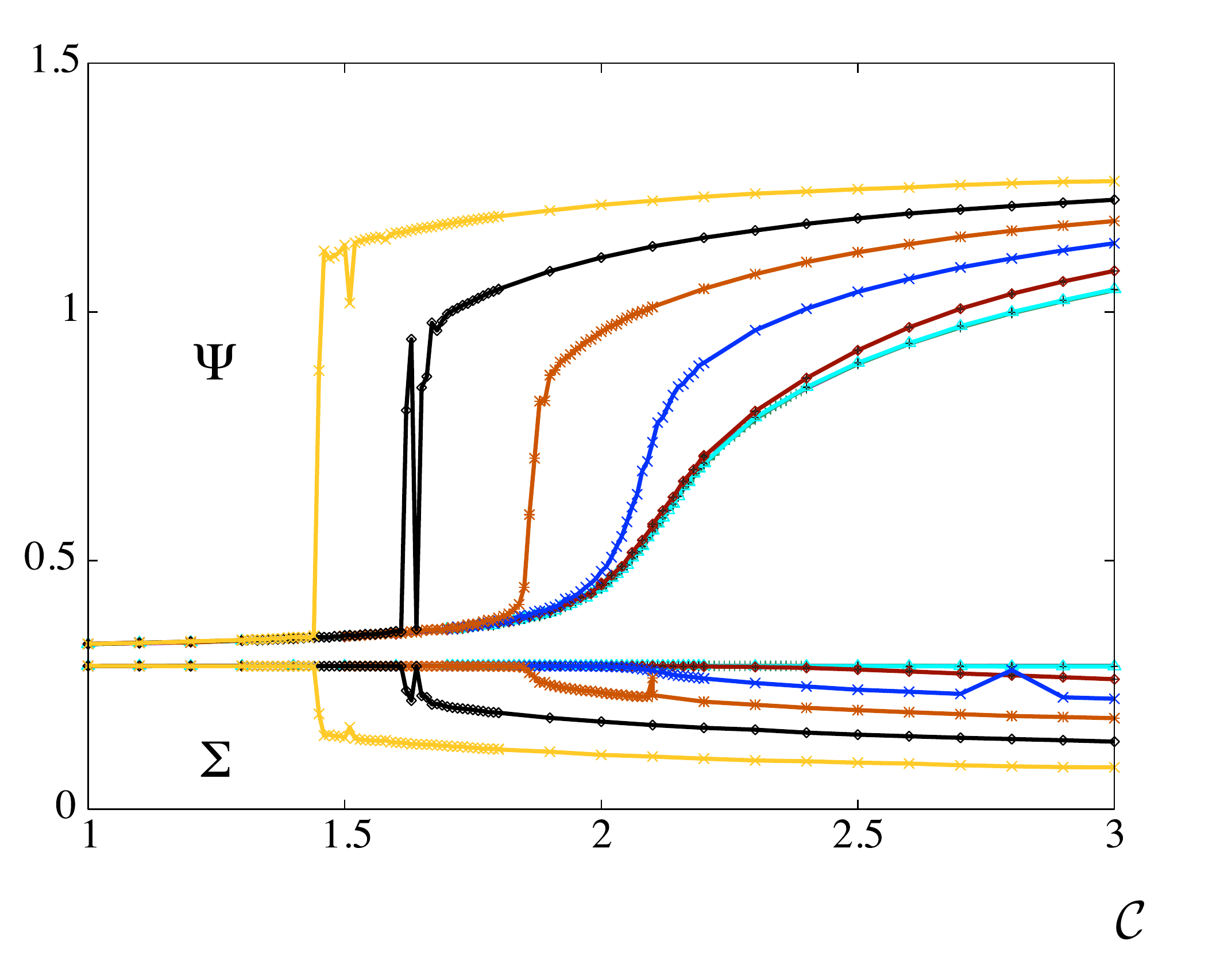}
\caption{Lower set of curves: order parameter $\Sigma$ as a function of the control parameter $\mathcal{C}$ and (from left to right in the figure) the same values of the interaction length $\sigma$ used in Fig.\ref{fig:psi}. The curves for $\sigma=0.3,0.4,0.5$ overlap with the line $\Sigma=1/\sqrt{12}\approx 0.2887$, the root-mean-square of a uniform distribution in the $[0,1]$ interval. For comparison we have also reproduced (vertically shifted by an arbitrary amount) the different lines of Fig.\ref{fig:chi} to show that the transition from flocking to non-flocking in the location of the particles occurs at the same value as the transition from order to disorder in the velocities.} \label{fig:psisigma}
\end{figure}

Above the transition point $\mathcal{C}>\mathcal{C}^*(\sigma)$ the order parameter $\Psi$ is different from zero, indicating an {\bfseries ordered} (O) phase in which a large fraction of particles move preferentially on average in the same direction. For $\mathcal{C}<\mathcal{C}^*(\sigma)$, the system is in the {\bfseries disordered} (D) phase, where the different trajectories are uncorrelated and, on average, half of the particles move to the right and half to the left. It turns out that the ordered phase can appear in two forms: a {\slshape spatially homogeneous} (OH) phase (characterized again by a flat and time-independent spatial pdf) and a {\slshape clustered} (OC) phase in which a macroscopic fraction of particles cluster in a particular location of space that moves with global constant velocity. In the OC phase, there is flocking as a large fraction of particles cluster together in the same region of space and move with the same velocity in the same direction. This traveling cluster induces a moving density profile $\rho(x,t)=\rho(x\pm V_0t)$. In the non-clustering ordered scenario, the OH phase, the majority of particles move in the same direction. To be able to distinguish between the two possible OC and OH  ordered phases, we introduce a second order parameter that originates from the normalized root-mean-square $\Sigma=\sigma[x]/L$ of the spatial pdf $\rho(x,t)$:
 \begin{eqnarray}
 \sigma[x]=\sqrt{\overline{x^2}-\bar{x}^2},\quad \overline{x^n}=\int_0^Ldx\,x^n\rho(x,t).\label{eq:sigma_order_2}
 \end{eqnarray}
The order parameter is $\mathbf{\Sigma}=\left\langle \Sigma\right\rangle,$  where $\left\langle \Sigma\right\rangle$ denotes a time average in the steady state. If the pattern is homogeneous, the standard deviation is that of a flat distribution $\rho(x,t)=\dfrac{1}{L}, x\in[0,L]$ or $\Sigma=1/\sqrt{12}\approx 0.289$. For a single localized pattern\cite{footnote}, $\Sigma$ scales as the width of the pattern divided by $L$. As shown in Fig.~\ref{fig:psisigma} for sufficiently low $\sigma\lesssim 0.3$ the order parameter $\mathbf{\Sigma}$ signals a transition from a homogeneous to a clustered phase at the same transition point $\mathcal{C}^*(\sigma)$ as the order parameter $\Psi$ indicates the transition from disorder to order. For better evidence, in this figure we have plotted both order parameters $\mathbf{\Sigma}$ and $\Psi$.

\begin{figure}[ht]
\includegraphics[width =0.50\textwidth]{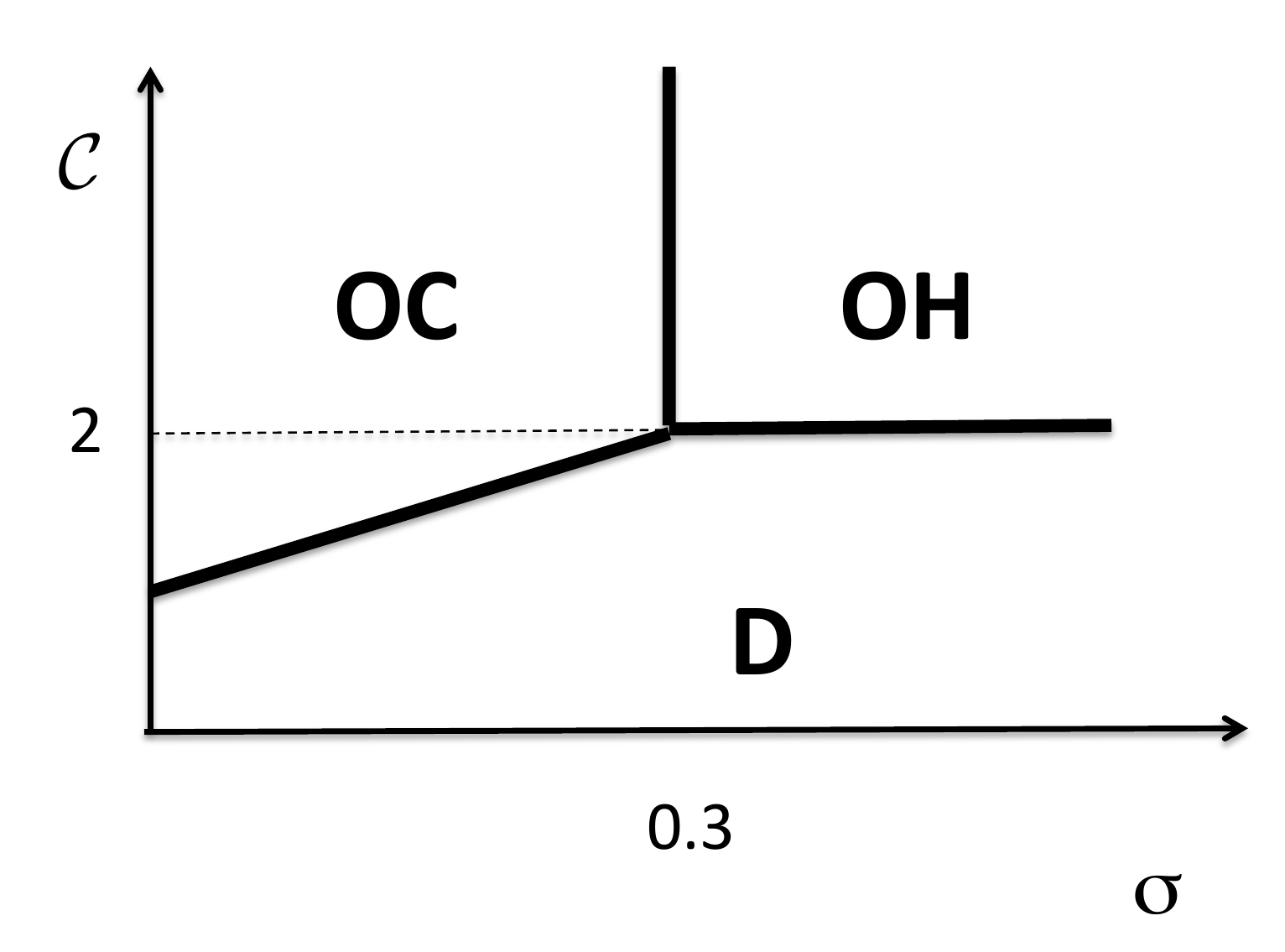}
\caption{Schematic (not to scale) phase diagram in the $(\sigma,{\cal C})$ space 
of parameters showing the different phases present in the steady state 
of the dynamical model discussed in the text. In the disordered (D) 
phase, particles move randomly and independently of each other to the 
right or to the left. In the ordered homogeneous (OH) phase, a large 
number of particles move synchronously in a preferred direction but are 
uniformly distributed in space. In the ordered clustered (OC) phase, 
particles, besides moving synchronously, stay close to each other in the 
same region of space.} \label{fig:phased}
\end{figure}

The phase diagram in the $(\sigma,\mathcal{C})$ space is schematized if Fig.\ref{fig:phased}. D is the disordered phase where particles have randomly distributed velocities and the density $\rho$ is uniform. In the OH (ordered homogeneous) phase, a majority of particles synchronize their velocities but the density of particles is still uniform. In the OC (ordered clustered) phase, particles cluster around a point in space that moves with velocity $+V_0$ or $-V_0$. 

\section{QUASI-ANALYTIC ESTIMATION OF THE SHAPE OF TRAVELING CLUSTERS}
\label{Sec6}

Let us consider a traveling solution of the advection-reaction equations (\ref{MF01}) and (\ref{MF02}). Without loss of generality, we will consider probability profiles that move to the right,
 \begin{equation}
\rho_{\pm}(x,t) = \rho_{\pm}(x - V_0 t). \label{traveling}
\end{equation}
Then, equations (\ref{MF01}) and (\ref{MF02}) take the form
\begin{align}
 - \lambda \left( \frac{\mathcal{C}}{\alpha}\nu_{\sigma} \left[\rho_{-} \right]\right)\rho_{+} 
+ \lambda \left( \frac{\mathcal{C}}{\alpha}\nu_{\sigma} \left[\rho_{+} \right]\right)\rho_{-} &= 0,
\label{MFtra1}\\
 - \lambda \left( \frac{\mathcal{C}}{\alpha}\nu_{\sigma} \left[\rho_{-} \right]\right)\rho_{+} 
+ \lambda \left( \frac{\mathcal{C}}{\alpha}\nu_{\sigma} \left[\rho_{+} \right]\right)\rho_{-} &= 2V_0\frac{\partial \rho_{-}}{\partial x}.
\label{MFtra2}
\end{align}
Equations (\ref{MFtra1}) and  (\ref{MFtra2}) imply
 \begin{equation*}
\frac{\partial \rho_{-}}{\partial x} = 0 ~~\Rightarrow~~  \rho_{-} = p_0,
\end{equation*}
where $p_0$ is a constant. 

Hence, using the model (\ref{lambmodelPOLI}) for the function $\lambda$ and after some algebraic manipulations, Eq.~(\ref{MFtra1}) can be rewritten in the form
 \begin{equation}
\mathcal{D}\rho_{+} = - \frac{\partial U(\rho_{+})}{\partial \rho_{+}}. \label{SoliEq}
\end{equation}
where
 \begin{equation}
U(\rho_{+}) = - \frac{2\alpha}{3\mathcal{C}\Gamma}\left( \Gamma\rho_{+} - 1\right)^{3/2} + \sigma\rho_{+}^{2},\label{PotentialEnergy}
\end{equation}
with
 \begin{equation*}
\Gamma = \frac{1}{p_0} + \left(\frac{2\mathcal{C}\sigma}{\alpha}\right)^2 p_0 = \frac{1}{p_0} + \left(N a\right)^2 p_0,
\end{equation*}
while the linear operator $\mathcal{D}$ has the form
 \begin{equation*}
\mathcal{D}\rho_{+} = \int_{x-\sigma}^{x+\sigma} \left(\rho_{+} (x^{\prime}) - \rho_{+} (x)\right) dx^{\prime}.
\end{equation*}
Note that this operator can be expanded, 
 \begin{equation*}
\mathcal{D} = \sum_{j = 1}^{\infty} \frac{2\sigma^{2j + 1}}{\left(2j + 1 \right)!} \frac{\partial^{2j} }{\partial x^{2j}}.
\end{equation*}

In order to give an analytic estimation for the density profile of the cluster, let us just take the first order in the expansion of the operator $\mathcal{D}$, that is,
 \begin{equation*}
\mathcal{D} \approx \frac{\sigma^{3}}{3} \frac{\partial^{2} }{\partial x^{2}}.
\end{equation*}
Then, Eq.~(\ref{SoliEq}) becomes  a Newton-type equation, which can be integrated,
\begin{equation*}
\frac{\partial \rho_{+}}{\partial x} =\sqrt{6(E-U(\rho_{+}))/\sigma^3} 
\end{equation*}
where $E$ is a conserved quantity, typically related to the energy in a mechanical problem. Then,
 \begin{equation}
\frac{\sqrt{6}(x - V_0 t)}{\sigma} = \int_{\rho_{0}}^{\rho_{+}} \frac{d\rho}{\sqrt{\left(E - U(\rho)\right)/\sigma}}, \label{integral}
\end{equation}
where $\rho_{0}$ denotes some initial condition. Since, the system is invariant under spatial translations and the solution is moving, the election of $\rho_{0}$ is not relevant.

The result of the integral in 
Eq.~(\ref{integral}) is a long expression which can not be analytically inverted. Therefore, the last step must be carried out numerically.

To perform our estimation of the shape of the cluster, we look for the homoclinic orbits of the Newton-type system. For a given value of the free parameter $p_0$, this fixes the value of the energy, say $E_H (p_0)$ at the homoclinic orbit. This energy is the same as the hyperbolic point that supports the solitary wave, that is $E_H (p_0) = U_h$, where $U_h$ is the potential-like function Eq.~(\ref{PotentialEnergy}) evaluated at the hyperbolic fixed point. 
From the numerical simulations, it seems that almost all the particles are absorbed by the traveling cluster. For small $p_0$, the hyperbolic point corresponds to $\rho_{+} = \rho_{-} = p_0$. We note that the limit $p_0 = 0$ is singular and does not admit a
solitary wave solution. However, for small $p_0$, and after normalization, we can obtain a good estimation of the cluster.
In other words, if $\rho_{+} = \Phi (x - V_0 t, p_0) $ corresponds to the homoclinic orbit of the Newton-type system for a given value of $p_0$, our analytic estimation for the density profile of the cluster corresponds to
\begin{align}
\rho_{+} &= \lim_{p_0\rightarrow 0} \frac{\Phi (x - V_0 t, p_0)}{\int_{0}^{L} \Phi (z, p_0)dz},
\label{Cluster1}\\
\rho_{-} &= 0.
\label{Cluster2}
\end{align}

\begin{figure}[ht]
\includegraphics[width =3.0 in]{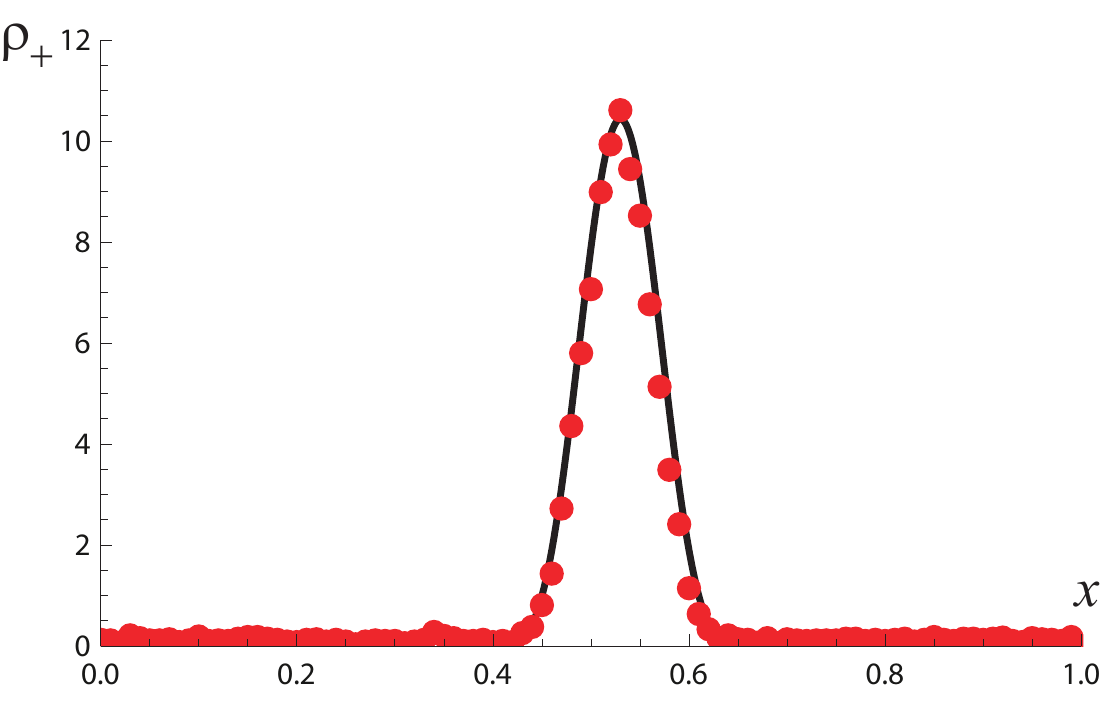}
\caption{Continuous curve: Mean field density profile for the cluster, as the result of inverting equation (\ref{integral}) for $L=1$, $\sigma=0.05$ and $\mathcal{C} = 1.7$. Dots: Data from direct numerical simulation of the microscopic rule, for the same parameters.} \label{figsoli}
\end{figure}

Figure \ref{figsoli} displays our result of inverting Eq.~(\ref{integral}), following the protocol described above. To estimate the limit in Eq. (\ref{Cluster1}), we have taken a small $p_0$ ($p_0 = 10^{-3}$ in Fig.~\ref{figsoli}), noting that after normalization, the result does not seem to be very sensitive to the value of $p_0$. The dots in figure \ref{figsoli} come from direct numerical simulation of the microscopic rule. As we see, the agreement between our spatially extended mean field theory, and the direct numerical simulation of the microscopic rule, is satisfyingly good.

\medskip

\section{Summary and final remarks}
\label{Sec7}

We have presented a model for active matter, which is based on interacting persistent random walkers in one dimension. The microscopic rule is time-continuous; therefore, any values of the active particles' speed have physical significance. Following a similar strategy as that  in~\cite{Escaff}, we are able to write a set of advection-reaction equations that describe the spatiotemporal evolution of the densities of particles in each state of motion (moving right or moving left). These equations correspond to a spatially extended mean-field theory. Hence we are neglecting the inherent fluctuations of the system. In order to check the prediction of this approximation, we have performed direct numerical simulations of the microscopic rule.

Our control parameter, Eq.~(\ref{ContPar}), measures both the coupling strength and the density of particles. Increasing the control parameter, 
we have observed a transition to flocking. The nature of this transition, however, strongly depends on the range of interaction $\sigma$. For large $\sigma$, the system behaves as predicted by the spaceless mean field theory. That is, for $\sigma^{*}<\sigma<L/2$, the system behavior is well predicted by the fully connected (or all-to-all interaction) case $\sigma=L/2$.
More precisely, in this region of large $\sigma$, the system exhibits a second order transition to a flocking state, which is characterized by a spatially uniform flux of particles. The critical value of the control parameter, for which the system exhibits this transition to homogeneous flocking, seems to be the same as that for the fully connected system, that is Eq.~(\ref{Ccritico}). In contrast, for $\sigma<\sigma^{*}$, the transition to flocking is characterized by the formation of a cluster. The transition is first order, and occurs for lower values of the control parameter than the one predicted by Eq. (\ref{Ccritico}).

It is possible to conjecture that sufficiently increasing the system system size, we might end up in the short range interaction regimen. Then, the transition to flocking should be first order and characterized by cluster formation. Note that, the advection-reaction system gives a good approximation of the density profile of the cluster. This noiseless nonlinear system seems thus to be a good candidate for analytic investigation of active matter. The model should of course be extended to two and three dimensions. For the time being, we leave this challenge to future work. 

\section*{acknowledgments}

DE thank FONDECYT project N¼ 1170669 for financial support.
RT acknowledges financial support from Ministerio de Econom\'ia y 
Competitividad (MINECO) and Fondo Europeo de Desarrollo Regional (FEDER) 
under project ESOTECOS FIS2015-63628-C2-1-R.

DE and RT acknowledge the warm hospitality at UCSD, where most of this work 
was carried out.

\medskip


\begin{thebibliography}{99}

\bibitem{Einstein} A. Einstein, {\emph{Investigations on the theory of the brownian movement}} (Dover Publications, INC., 1956)

\bibitem{Bausch} Schaller, V., Weber, C., Semmrich, C., Frey, E. and
Bausch, A. R. Nature 467, 73Ð77 (2010).

\bibitem{Dogic} Sanchez, T., Chen, D. T. N., DeCamp, S. J., Heymann, M. and Dogic, Z. Nature 491, 431Ð434 (2012).

\bibitem{Bartolo} Bricard, A., Caussin, J.-B., Desreumaux, N., Dauchot, O. and Bartolo, D. Nature 503, 95Ð98 (2013).

\bibitem{Kaiser} D. Kaiser, Nat. Rev. Microbiol. 1, 45 (2003).

\bibitem{Kasyap} T. V. Kasyap, Donald L. Koch, and Mingming Wu, Physics of Fluids 26, 081901 (2014).

\bibitem{Parrish} J.K. Parrish and L. Edelstein-Keshet, Science 248, 99 (1999).

\bibitem{Popkin} Gabriel Popkin, Nature, 529, 16 (2016). 


\bibitem{Vicsek} Vicsek, T., Czir—k, A., Ben-Jacob, E., Cohen, I. and Shochet, O. Phys. Rev. Lett. 75, 1226Ð1229 (1995).

\bibitem{Toner} Toner, J. and Tu, Y. Phys. Rev. Lett. 75, 4326Ð4329 (1995).

\bibitem{Toner2} Toner, J. and Tu, Y. Phys. Rev. E 58, 4828Ð44858 (1998).

\bibitem{Vicsek2} Czir—k, A., Barab\'asis A-L, I. and Vicsek, T. Phys. Rev. Lett. 82, 209Ð212 (1999).


\bibitem{Chate1} Gr\'egoire G., and Chat\'e H. Phys. Rev. Lett. 92, 025702 (2004).

\bibitem{Vicsek3} Nagy, M., Daruka I., and Vicsek, T. Physica A 373, 445 (2007).

\bibitem{Aldana} M. Aldana, V. Dossetti, C. Huepe, V. M. Kenkre, and H. Larralde. Phys. Rev. Lett. 98, 095702 (2007).


\bibitem{Chate2} Hugues ChatŽ, Francisco Ginelli, Guillaume GrŽgoire, and Franck Raynaud. Phys. Rev. E 77, 046113 (2008).

\bibitem{Peruani1} Fernando Peruani, Andreas Deutsch, and Markus Bar, Phys. Rev. E 74, 030904(R) (2006).

\bibitem{Marchetti} Aparna Baskaran and Cristina Marchetti. Phys. Rev. Lett. 101, 268101 (2008).

\bibitem{Peruani2} Francesco Ginelli, Fernando Peruani, Markus Bar, and Hugues Chat\'e.  Phys. Rev. Lett. 104, 184502 (2010).

\bibitem{Solon} A. P. Solon and J. Tailleur Phys. Rev. Lett. 111, 078101 (2013). 

\bibitem{Kac} Mark Kac, Rocky Mountain Journal Of Mathematics 4, 497 (1974). 

\bibitem{Katja1} Jaume Masoliver, Katja Lindenberg, and George H. Weiss, Physica A 157 891-898 (1989).

\bibitem{Masoliver} Jaume Masoliver, and George H. Weiss, Phys. Rev. E 49, 3852 (1994).

\bibitem{Katja2} Jaume Masoliver, and Katja Lindenberg, Eur. Phys. J. B 90, 107 (2017).



\bibitem{PRW} S. Goldstein. On diffusion by discontinuous movements, and on the telegraph equation. Quart. J. Mech. Appl. Math., 4:129Ð156, 1951;
Eric Renshaw and Robin Henderson. The correlated random walk. J. Appl. Probab., 18(2):403Ð414, 1981; 
George H. Weiss. Aspects and applications of the random walk. Random Materials and Processes. North-Holland Publishing Co., Amsterdam, 1994;
Eugene C. Eckstein, Jerome A. Goldstein, and Mark Leggas. The mathematics of suspensions:
Kac walks and asymptotic analyticity. In Proceedings of the Fourth Mississippi State Conference on Difference Equations and Computational Simulations (1999), volume 3 of Electron; 
George H. Weiss. Some applications of persistent random walks and the telegrapherÕs equation. Phys. A, 311(3-4):381Ð410, 2002;
Jaume Masoliver, Phys. Rev. E 96, 022101 (2017).




\bibitem{Burkholder} Eric W. Burkholder and John F. Brady, Phys. Rev. E 95, 052605 (2017).

\bibitem{Kirman} A. Kirman, Quart. J. Econ. 108, 137 (1993).

\bibitem{Raul} A. Fernandez-Peralta, R. Toral, A. Carro and M. San Miguel,  arXiv:1803.06861.

\bibitem{Pinto} I. L. D. Pinto, D. Escaff, U. Harbola, A. Rosas,
and K. Lindenberg, Phys. Rev. E 89, 052143
(2014).

\bibitem{Rosas1} A. Rosas, D. Escaff, I. L. D. Pinto, and K. Lindenberg, J. Phys.
A 49, 095001 (2016).

\bibitem{Rosas2} A. Rosas, D. Escaff, I. L. D. Pinto, and K. Lindenberg, Phys. Rev. E 95, 032104 (2017).

\bibitem{Escaff} Daniel Escaff, ItaloÕIvo Lima Dias Pinto, and Katja Lindenberg, Phys. Rev. E 90, 052111 (2014).


\bibitem{Turing} A. M. Turing, Philos. Trans. R. Soc. B 237, 37 (1952).

\bibitem{Fuentes} M. A. Fuentes, M. N. Kuperman, and V. M. Kenkre, Phys. Rev.
Lett. 91, 158104 (2003).

\bibitem{Emilio1} E. Hernandez-Garcia and C. Lopez,
Phys. Rev. E 70, 016216 (2004);

\bibitem{Emilio2} E. Heinsalu, E. Hernandez-Garcia, and C. Lopez, Phys. Rev. E
85, 041105 (2012).

\bibitem{Escaff2} M. G. Clerc, D. Escaff, and V. M. Kenkre, Phys. Rev. E 72,
056217 (2005).

\bibitem{Escaff3} M. G. Clerc, D. Escaff, and V. M. Kenkre, Phys. Rev. E
82, 036210 (2010).

\bibitem{Lejeune} O. Lejeune and M. Tlidi, J. Veg. Sci. 10, 201 (1999).

\bibitem{Escaff4} D. Escaff, C. Fernandez-Oto, M. G. Clerc,3 and M. Tlidi, Phys. Rev. E 91, 022924 (2015).

\bibitem{Marchetti2} S. Mishra, A. Baskaran, and M. C. Marchetti, Phys. Rev. E 81, 061916 (2010).

\bibitem{Marchetti3} A. Gopinath, M. F. Hagan, M. C. Marchetti, and A. Baskaran, Phys. Rev. E 85, 061903 (2012).

\bibitem{Ihle} T. Ihle, Phys. Rev. E 83, 030901 (2011);

\bibitem{Deutsch} H.-P. Deutsch, J. Stat. Phys. 67,1039 (1992). 
 
\bibitem{footnote}If there were more than one localized pattern in the system (say two solitary waves) then one has to be more careful in the definition of this order parameter, but we have not found these states for the range of parameters considered in our simulations.

\bibitem{morework} A more detailed account of the influence of the density of particles $\ell=L/N$ is outside the scope of this paper and will be published subsequently.

\end{thebibliography}

\end{document}